\documentclass[12pt]{article}
 
\usepackage{amsfonts, amsthm}
\usepackage{amssymb}
\usepackage{amsmath}  
\usepackage{graphicx, comment, color, hyperref}
\usepackage{authblk}
\usepackage{mathrsfs}

\usepackage[a4paper]{geometry}
\newtheorem{theorem}{Theorem}

\theoremstyle{remark}
\newtheorem{remark}{Remark}

\newcommand{\tr}{\text{\rm tr}}
\newcommand{\E}{\mathbb{E}}

\newcommand{\bfy}{\mathbf{y}}

\title{Quantum Kernel Method in the Presence of Noise}

\author{Salman Beigi}
\affil{\small \it QuOne Lab, Phanous Research and Innovation Centre, Tehran, Iran}

\begin{document}

\maketitle

\begin{abstract}
Kernel method in machine learning consists of encoding input data into a vector in a Hilbert space called the \emph{feature space} and modeling the target function as a linear map on the feature space. Given a cost function, computing such an optimal linear map requires computation of a \emph{kernel matrix} whose entries equal the inner products of feature vectors. In the \emph{quantum kernel method} it is assumed that the feature vectors are quantum states in which case the quantum kernel matrix is given in terms of the overlap of quantum states. In practice, to estimate entries of the quantum kernel matrix one should apply, e.g., the SWAP-test and the number of such SWAP-tests is a relevant parameter in evaluating the performance of the quantum kernel method. Moreover, quantum systems are subject to noise, so the quantum states as feature vectors cannot be prepared exactly and this is another source of error in the computation of the quantum kernel matrix. Taking both the above considerations into account, we prove a bound on the performance (generalization error) of the quantum kernel method.

\end{abstract}

\section{Introduction}

Quantum kernel method is one of the main proposals for achieving quantum advantages in machine learning~\cite{SK19, HCTHKCG19}. In this method, a data point $x$ is encoded into a quantum state $\rho_x$ and its associated value is given by $f(x)=\tr\big(\rho_xO\big)$, where $O$ is an (unknown) quantum observable with $\|O\|\leq 1$. The goal is to find a \emph{hypothesis function} $h(x)$ that minimizes
\begin{align}\label{eq:gen-error}
\E_x\Big[\big|h(x) - \tr\big(\rho_xO\big)\big|\Big],
\end{align}
where the expectation is computed with respect to some fixed distribution on the space of data points $x$. It is shown in~\cite{Huang+21} that given access to a \emph{training dataset} $\{(x_i, y_i):\, i=1,\dots, N\}$ with $y_i = f(x_i)$, where $x_i$'s are drawn independently from the predetermined distribution, we can find a function $h(x)$ such that with high probability, the error~\eqref{eq:gen-error} is upper bounded by $\mathcal O(\sqrt{\mathbf y^T K^{-1}\mathbf y/N})$.\footnote{See below for the more precise bound.} Here, $\mathbf y=(y_1, \dots, y_N)^T$ and $K$ is the $N\times N$ matrix with entries $K_{ij}= \tr\big(\rho_{x_i}\rho_{x_j}\big)$, called the \emph{kernel matrix}. The idea in~\cite{Huang+21} is to model $h(x)=h_\Theta(x)$ as\footnote{Note that $\|O\|\leq 1$ and $f(x)=\tr(\rho_x O)$ belongs to the interval $[-1, 1]$. This is why in $h_\Theta(x)$ the value of $\tr(\rho_x \Theta)$ is mapped to the interval $[-1, 1]$.} 
$$h_\Theta(x)= \min\{1, ~ \max\{-1, \tr(\rho_x \Theta)\}  \},$$
where $\Theta$ is a \emph{weight matrix} to be found. Then, $\Theta$ is set as the solution of the optimization problem
$$\min_\Theta \sum_{i=1}^N  \Big(  \tr(\rho_{x_i}\Theta) - y_i  \Big)^2  + \lambda \tr(\Theta^\dagger \Theta),$$
where $\lambda\geq 0$ is a \emph{regularization parameter} that is fixed in advance. It is not hard to verify that this optimization problem has the explicit solution
$$\Theta= \sum_{i,j=1}^N (K+\lambda I)^{-1}_{ij} y_j \rho_{x_j}.$$
It is shown in~\cite{Huang+21} that this choice of $\Theta$ results in the aforementioned upper bound on~\eqref{eq:gen-error}.

While fixing the function $f(x)=\tr(\rho_x O)$, we may apply the above ideas even considering other feature maps and kernel matrices. Suppose that we have another feature map $x\mapsto \phi_x$ in which $\phi_x$ is an arbitrary vector in some Hilbert space beyond the quantum Hilbert space. In this case, our function $h_{\Theta}(x)$ would be of the form $h_\Theta(x)= \min\{1, ~ \max\{-1, \langle \phi_x , \Theta\rangle\}  \}$, where now $\Theta$ is a vector in the same Hilbert space. Then, the relevant kernel matrix is given by $W_{ij} = \langle \phi_{x_i}, \phi_{x_j}\rangle$. Interestingly, the result of~\cite{Huang+21} holds even considering such generalized kernels beyond the quantum kernel.

\begin{theorem}\label{thm:Huang+21} \emph{\cite{Huang+21}} 
Let $f(x) = \tr(\rho_x O)$ where $x\mapsto \rho_x$ is some encoding map and $O$ is an (unknown) observable with $\|O\|\leq 1$. Suppose that we are given a training dataset $\{(x_i, y_i):\, i=1,\dots, N\}$ with $y_i = f(x_i)=\tr\big(\rho_{x_i}O\big)$, where $x_i$'s are drawn independently at random. Let $x\mapsto \phi_x$, for any data point $x$, be a \emph{feature map} where $\phi_x$ is some vector in a Hilbert space. Let $W$ be the associated kernel matrix with entries $W_{ij} = \langle \phi_{x_i}, \phi_{x_j}\rangle$. Also, let $\lambda\geq 0$ be a regularization parameter.   
Then, we can find a function $h(x)$ such that with probability at least $1-\delta$ over the choice of the training dataset, we have
$$\E_x\Big[\big|h(x) - \tr\big(\rho_xO\big)\big|\Big] \leq \mathcal O\left( \sqrt{\frac{\lambda^2\bfy^T (W+\lambda I)^{-2}\bfy}{N}} +\sqrt{\frac{\bfy^T (W+\lambda I)^{-1} W(W+\lambda I)^{-1}\bfy}{N}}+\sqrt{\frac{\log 1/\delta}{N}} \right),$$
where $\bfy = (y_1, \dots, y_N)^T$.
\end{theorem}

\begin{remark}\label{rem:thm-Huang+21}
The bound of Theorem~\ref{thm:Huang+21} can be simplified as follows. First, using $\sqrt x+\sqrt y\leq 2\sqrt{x+y}$ we have 
\begin{align*}
&\mathcal O\left( \sqrt{\frac{\lambda^2\bfy^T (W+\lambda I)^{-2}\bfy}{N}} +\sqrt{\frac{\bfy^T (W+\lambda I)^{-1} W(W+\lambda I)^{-1}\bfy}{N}} \right) \\
&\qquad\qquad  \leq \mathcal O\left( \sqrt{\frac{\lambda^2\bfy^T (W+\lambda I)^{-2}\bfy +\bfy^T (W+\lambda I)^{-1} W(W+\lambda I)^{-1}\bfy}{N}} \right)\\
&\qquad\qquad  = \mathcal O\left( \sqrt{\frac{\bfy^T (W+\lambda I)^{-1} (W+\lambda^2 I)(W+\lambda I)^{-1}\bfy}{N}} \right).
\end{align*}
Next, we note that if $\lambda\leq 1$, then $W+\lambda^2\preceq W+\lambda I$, and if $\lambda>1$, 
$$W+\lambda^2 I\preceq \lambda W + \lambda^2= \lambda(W+\lambda I).$$
Therefore, letting $\lambda'=\max\{1, \lambda\}$, in both cases we have $W+\lambda^2 I\preceq \lambda'(W+\lambda I)$. Using this in the above equation, the bound of Theorem~\ref{thm:Huang+21} can be simplified to
$$\E_x\Big[\big|h(x) - \tr\big(\rho_xO\big)\big|\Big] \leq \mathcal O\left( \sqrt{\frac{\lambda'\bfy^T (W+\lambda I)^{-1} \bfy}{N}}+\sqrt{\frac{\log 1/\delta}{N}} \right),$$
where $\lambda'=\max\{1, \lambda\}$.
\end{remark}

\section{Our contribution}
There are two main assumptions behind Theorem~\ref{thm:Huang+21}. First, it is assumed that the encoding process is noiseless and given $x$, the quantum state $\rho_x$ can be generated exactly. Second, we can compute entries of the kernel matrix $K_{ij}= \tr\big(\rho_{x_i}\rho_{x_j}\big)$ exactly. Nevertheless, as will be explained none of these two assumptions are realistic. 

Considering the quantum kernel matrix associated to the feature map $x\mapsto \phi_x=\rho_x$, in order to find $h(x)$ using the algorithm suggested by Theorem~\ref{thm:Huang+21}, we need to compute entries of the kernel matrix given by $K_{ij} = \tr(\rho_{x_i}\rho_{x_j})$. This can be done by applying the SWAP-test (or its variants) on $\rho_{x_i}, \rho_{x_j}$. However, in practice, the process of generating $\rho_{x_i}, \rho_{x_j}$ is subject to noise, e.g., depolarizing noise, in which case instead of $\rho_x$ we obtain its noisy version $\tilde \rho_x$ given by
\begin{align}\label{eq:rho-tilde}
\tilde \rho_x = (1-p)\rho_x + p \frac{I}{D},
\end{align}
where $D$ is the dimension of the quantum state $\rho_x$. In this case, we do not (at least directly) have access to the kernel matrix $K$, but a noisy version of it given by
$$\widetilde K_{ij} = \tr\big(\tilde \rho_{x_i}\tilde \rho_{x_j}\big).$$
Moreover, as mentioned above to compute entries of the kernel matrix, we apply SWAP-test. However, note that the output of the SWAP-test is just a Bernoulli random variable whose \emph{expectation} is the desired value. Thus, to obtain a good approximation we need to repeat the SWAP-test many times, say $m$ times, and take the average of those $m$ outcomes as our estimator. We note that by increasing $m$ we obtain a more accurate estimation. Nevertheless, this process itself is another source of error in estimating entries of the kernel matrix.

To summarize, in practice, both the noise parameter $p$ in~\eqref{eq:rho-tilde} and the number of repetitions $m$ of the SWAP-test in estimating any entry of the kernel matrix, affect the performance of the quantum kernel method. 
 
The above considerations have been addressed in~\cite{Wang+21} where Theorem~\ref{thm:Huang+21} is generalized in the presence of noise and with the complications in estimating kernel entries. It is concluded in~\cite{Wang+21} that ``\emph{even though $m$ is set as sufficiently large, the generalization error bound can still be very large induced by $p$,}" which is very unintuitive. In this paper we improve on the error analysis of~\cite{Wang+21} and show that assuming that $m$ is sufficiently large, the result of~\cite{Huang+21} still holds true with a small modification. 

Here is the main result of this paper.

\begin{theorem}\label{thm:main}
Let $f(x)=\tr\big(\rho_x O\big)$ where $x\mapsto \rho_x$ is some encoding map and $O$ is an (unknown) observable with $\|O\|\leq 1$. Suppose that we are given dataset $\{(x_i, y_i):\, i=1,\dots, N\}$ with $y_i=f(x_i)$, and for each $i$ can generate noisy states $\tilde \rho_{x_i}$ given by
\begin{align}\label{eq:tilde-rho}
\tilde \rho_{x_i} = (1-p) \rho_{x_i} + p \frac{I}{D},
\end{align}
where $0\leq p<1$ is some noise parameter and $D$ is the dimension of the underlying quantum system. Suppose that to estimate $\tr\big(\tilde \rho_{x_{i}}\tilde \rho_{x_j}\big)$, for each $i,j$, we perform $m$ SWAP-tests on independent copies of $\tilde \rho_{x_i}, \tilde \rho_{x_j}$. Then, assuming that 
$$m\geq \frac{8N}{\lambda^2}\log (2N/\delta),$$ 
we can find a function $h(x)$ such that with probability at least $1-\delta$ over the choice of the training dataset and the outcomes of the SWAP-tests, we have
$$\E_x\Big[\big|h(x) - \tr\big(\rho_xO\big)\big|\Big] \leq \mathcal O\left( \sqrt{\frac{\lambda' \bfy^T (K+\lambda I)^{-1}\bfy}{(1-p)^2N}} +\sqrt{\frac{\log 1/\delta}{N}} \right),$$
where $\bfy = (y_1, \dots, y_N)^T$ and $K$ is the kernel matrix given by $K_{ij} = \tr\big(\rho_{x_i}\rho_{x_j}\big)$, $\lambda>0 $ is a regularization parameter, and $\lambda'=\max\{1, \lambda\}$.

\end{theorem}

This theorem shows that, fixing the other parameters, as long as $m$ is larger than $cN\log N$, for some constant $c$, we can find a hypothesis $h(x)$ with a small generalization error. This is an improvement over~\cite{Wang+21} and~\cite{Petersetal} that recommend $m$ to be of order of $N^3$ and $N^2$, respectively. Moreover, as our intuition suggests, larger choices for $m$ result in smaller generalization bounds even if the noise parameter $p$ is positive. Finally, in this theorem we consider only the depolarizing noise. Nevertheless, as the proof of the theorem given in the following section suggests, similar results can be derived for other noise models.

\section{Proof of Theorem~\ref{thm:main}}

Let $\widetilde K$ be the kernel matrix associated with $\tilde \rho_{x_i}$'s:
$$\widetilde K_{ij} = \tr\big( \tilde \rho_{x_i}\tilde \rho_{x_j} \big).$$
Using~\eqref{eq:tilde-rho} it is not hard to verify that 
\begin{align}\label{eq:tilde-K}
\widetilde K = (1-q) K + q \frac{I}{D},
\end{align}
where $q= 1-(1-p)^2\leq 1$.

Let $R_{ij}^{k}$, $1\leq k\leq m$, be the Bernoulli random variable associated to the outcome of the $k$-th SWAP-test on $\tilde \rho_{x_i}, \tilde \rho_{x_j}$. Then, our estimation of the kernel matrix denoted by $\widehat K$ is given by
$$\widehat K_{ij} =  \frac{1}{m}\Big(\sum_{k=1}^m R_{ij}^{(k)}\Big) E_{ij}, \qquad i\neq j,$$
where $E_{ij}=E_{ji}$ is the matrix all of whose entries are zero except the $ij$-th and $ji$-th entries which are equal to $1$. 
We note that $\widehat K$ is a random matrix that is hermitian and 
$$\E[\widehat K]= \widetilde K.$$ 
We claim that with high probability $\widehat K$ is close to $\widetilde K$. To this end, we use the \emph{matrix Hoeffding bound}.

\begin{theorem} \emph{\cite{MJCFT12} (Matrix Hoeffding bound)}
Let $X_1, \dots, X_m$ be independent random hermitian matrices of size $N\times N$, and let $A_1, \dots, A_m$ be hermitian matrices satisfying\footnote{$X\succeq Y$ means that $X-Y$ is positive semidefinite.}
$$\mathbb E[X_j]=0, \quad X_j^2\preceq A^2_j, \qquad \forall j.$$
Then, we have
$$\Pr\Big[  \sum_j X_j \npreceq tI  \Big]\leq N e^{-t^2/2\sigma^2},$$
where $\sigma^2 = \|\sum_j A_j^2\|$.
\end{theorem}

For any $i,j$ and $1\leq k\leq m$ let
$$X_{ij}^{(k)} := \frac 1m \big(\widetilde K_{ij}- \widehat R_{ij}^{(k)}  \big) E_{ij}.$$
Then, by the above discussion we have $\E[X_{ij}^{(k)}] =0$. Moreover, $\big|\widetilde K_{ij}- \widehat R_{ij}^{(k)}  \big|\leq 1$ yields
$$\Big(X_{ij}^{(k)}\Big)^2 \preceq \frac {1}{m^2}\Big(E_{ii} + E_{jj}\Big).$$
We also have 
$$\sigma^2= \Big\|\sum_{i,j, k}\frac {1}{m^2}\big(E_{ii} + E_{jj}\big)\Big\| = \frac{N}{m}.$$
Therefore, by the matrix Hoeffding inequality for $t=\lambda/2$,
with probability at least $1-N e^{-t^2m/2N}=1-N e^{-\lambda^2m/8N}$ we have 
\begin{align}\label{eq:Ktlmabda}
\widehat K+\frac \lambda 2 I\succeq \widetilde K\succeq 0.
\end{align}
This means that with high probability, $\widehat K + \frac{\lambda}{2}I$ is positive semidefinite, and can be considered as a kernel matrix. Using this kernel matrix in Theorem~\ref{thm:Huang+21} with the regularization parameter $\lambda/2$, and following Remark~\ref{rem:thm-Huang+21}, we can find a function $h(x)$ such that with probability at least $1-\delta' - N e^{-\lambda^2m/8N}$,
$$\E_x\Big[\big|h(x) - \tr\big(\rho_xO\big)\big|\Big] \leq O\left( \sqrt{\frac{\lambda'\bfy^T (\widehat K+\lambda I)^{-1} \bfy}{N}}+\sqrt{\frac{\log 1/\delta'}{N}} \right).$$
We note that by assumption $m\geq \frac{8N}{\lambda^2}\log (2N/\delta)$ which implies $N e^{-\lambda^2m/8N}\leq \delta/2$. Therefore, letting $\delta'=\delta/2$, with probability at least $1-\delta$ we have 
\begin{align}\label{eq:bound-intermed}
\E_x\Big[\big|h(x) - \tr\big(\rho_xO\big)\big|\Big] \leq O\left( \sqrt{\frac{\lambda'\bfy^T (\widehat K+\lambda I)^{-1} \bfy}{N}}+\sqrt{\frac{\log 1/\delta}{N}} \right).
\end{align}
Then, to prove the theorem it suffices to show that $(\widehat K+\lambda I)^{-1}\preceq \frac{2}{1-q}(K+\lambda I)^{-1}$. 

First, note that~\eqref{eq:Ktlmabda} implies $\widehat K +\lambda I\succeq \widetilde K+\frac \lambda 2 I$. Then, by the operator monotonicity of $x\mapsto 1/x$, we have 
$$(\widehat K +\lambda I)^{-1}\preceq \Big(\widetilde K+ \frac \lambda 2 I\Big)^{-1} \preceq \Big(\frac 12\widetilde K+\frac \lambda 2 I\Big)^{-1} = 2\big(\widetilde K+\lambda I\big)^{-1}.$$
Next, using~\eqref{eq:tilde-K} we find that 
$$(\widehat K +\lambda I)^{-1}\preceq 2 \Big( (1-q)K+(\lambda+q/D)I\Big)^{-1} \preceq  2 \Big( (1-q)K+(1-q)\lambda I\Big)^{-1} = \frac{2}{1-q} ( K +\lambda I)^{-1}.$$
Using this in~\eqref{eq:bound-intermed} the desired bound is obtained. 

\hfill$\Box$

\begin{remark}
We note that having $\widehat K$ as an estimation of $\widetilde K$, by~\eqref{eq:tilde-K} we can find an estimation of $K$ if $p$ is known. In practice, this requires quantum tomography in order to find an estimation of $p$, which itself is subject to error. The proposed algorithm in the above proof and its analysis, however, is ignorant of the exact value of $p$; it is the generalization bound that depends on $p$ but not the algorithm itself.
\end{remark}

{~~}

\end{document}